%% file: ecfaTemplate.tex
\documentclass[11pt,a4paper]{scrartcl}

\usepackage{ECFAreport}

\usepackage{ECFAreport_definitions}

\usepackage{xpatch}
\makeatletter
\xpatchcmd\@HepConStyle
 {\edef\@upcode{\updefault}}
 {\ifdefined\shapedefault\edef\@upcode{\shapedefault}\else\edef\@upcode{\updefault}\fi}
 {}{}
\makeatother

\RedeclareSectionCommand[
  beforeskip=1.\baselineskip,
  afterskip=.5\baselineskip]{subsection}
\RedeclareSectionCommand[
  beforeskip=1.\baselineskip,
  afterskip=.5\baselineskip]{subsubsection}

\input{preamble}

\DeclareCiteCommand{\citejournal}[\mkbibbrackets]
  {\usebibmacro{prenote}}
  {\usebibmacro{citeindex}%
   \printtext[bibhyperref]{\printfield{journaltitle}}%
   \iffieldundef{volume}
     {}%
     {\setunit{\addspace}%
     \printtext[bibhyperref]{\printfield{volume}}}%
   \setunit{\addspace}%
   \printtext[bibhyperref]{(\printdate)}%
   \iffieldundef{pages}
     {}
     {\setunit{\addspace}%
     \printtext[bibhyperref]{\printfield{pages}}%
     }%
     }
  {\multicitedelim}
  {\usebibmacro{postnote}}
  
\DeclareCiteCommand{\citesubmit}[\mkbibbrackets]
  {\usebibmacro{prenote}}
  {\usebibmacro{citeindex}%
   \printtext[bibhyperref]{\printfield{journaltitle}}%
   \setunit{\addspace}%
   \printtext[bibhyperref]{(\printdate)}}
  {\multicitedelim}
  {\usebibmacro{postnote}}
  
  \DeclareCiteCommand{\citeconf}[\mkbibbrackets]
  {\usebibmacro{prenote}}
  {\usebibmacro{citeindex}%
   \printtext[bibhyperref]{\printfield{howpublished}}%
   \setunit{\addspace}%
   \printtext[bibhyperref]{(\printdate)}}
  {\multicitedelim}
  {\usebibmacro{postnote}}

\title{Bottom quark forward-backward asymmetry at the future electron-positron collider FCC-ee}

\nolicence 

\notitlestamp 

\date{\today}

\addauthor{Marina Cobal}{\institute{1}\institute{2}}

\addauthor{Giovanni Guerrieri}{\institute{1}\institute{3}}

\addauthor{Hamzeh Khanpour}{\institute{1}\institute{4}\institute{5}\institute{6}}

\addauthor{Giancarlo Panizzo}{\institute{1}\institute{2}}

\addauthor{Michele Pinamonti}{\institute{1}\institute{2}} 

\addauthor{Laura Pintucci}{\institute{1}\institute{3}} 

\addauthor{Leonardo Toffolin}{\institute{1}\institute{3}} 

\addinstitute{1}{INFN Sezione di Trieste, Gruppo Collegato di Udine, Italy}
\addinstitute{2}{Dipartimento Politecnico di Ingegneria e Architettura, University of Udine, Via delle Scienze 206, 33100 Udine, Italy}
\addinstitute{3}{Physics Department, University of Trieste, Via Valerio 2, 34127 Trieste, Italy}
\addinstitute{4}{The Abdus Salam International Centre for Theoretical Physics (ICTP), Strada Costiera 11, 34151 Trieste, Italy}
\addinstitute{5}{School of Particles and Accelerators, Institute for Research in Fundamental Sciences (IPM), P.O.Box 19395-5531, Tehran, Iran}
\addinstitute{6}{Department of Physics, University of Science and Technology of Mazandaran, P.O.Box 48518-78195, Behshahr, Iran}

\abstract{\noindent

The Standard Model (SM) prediction for the \PZ-boson pole $\PQb$-quark 
forward-backward (FB) asymmetry is: $(A_{FB}^{0,b})_{th} = 0.1030 \pm 0.0002$. 
The LEP electron-positron collider measured instead $A_{FB}^{0,b} = 0.0992 \pm 0.0016$, value which presents the largest discrepancy with any of the SM predictions as of today. 
All the $A_{FB}^b$ measurements performed at LEP suffered however of an important statistical uncertainty and of different sources of systematic uncertainties.
This study shows that the proposed high-luminosity electron-positron collider FCC-ee, 
collecting orders of magnitude more data at the \PZ-pole than LEP, will 
significantly reduce the statistical uncertainties on the measurement of $A_{FB}^{0,b}$, thus allowing us to shed further light on this tension.}

\graphicspath{ {./logos/}{./figures/} }

\DeclareTOCStyleEntry[beforeskip=.2cm]{section}{section}

\begin{document}

\titlepage
\pagenumbering{arabic}\setcounter{page}{2}


\input{./sections/introduction/introduction.tex}

\input{sections/mainbody/everything}


\section{References}
\printbibliography[heading=none]

\end{document}

%% file: preamble.tex
\usepackage{todonotes}
\usepackage{xpatch}
\makeatletter
\xpretocmd{\todo}{\@bsphack}{}{}
\xapptocmd{\todo}{\@esphack}{}{}
\makeatother

\usepackage{booktabs}
\usepackage{csquotes}

\usepackage{mathtools}

\newcommand{\AFBobs}{A_{FB}^{b, \text{obs}}}

%% file: sections/introduction/introduction.tex
\section{Introduction}\label{sec:introduction}
%

The data accumulated by the large electron-positron (LEP) collider at CERN in the years between 1990 and 2000 was used to determine \PZ-boson properties with unprecedented high precision~\cite{ALEPH:2005ab}, providing an outstanding stress test of the SM. Among these results, the measurement of the forward-backwards asymmetry of $\PQb$-quarks ($A_{FB}^{0,b}$) produced in electron-positron collisions via $\Pep  \Pem \to \PZ \to  \PQb \PAQb$ allowed to determine the effective weak mixing angle $\sin^2\theta_W$ with high precision~\cite{ALEPH:2005ab,dEnterria:2020cgt,Altarelli:1989hv}. 
This asymmetry is created by the parity violation of the neutral weak current and it can be extracted from the differential cross-section of $\sigma(\Pep  \Pem \to \PZ \to  \PQb \PAQb)$, as in \cref{eq:fit_to_cos}~\cite{Tenchini:2008zz}, 

\begin{equation}
\frac{d\sigma}{d\cos {\theta}} = 
\sigma^{\text{tot}}_{b \bar {b}}
\left[\frac{3}{8}(1 + \cos^{2} \theta)+ 
A_{FB}^{b} \cos {\theta} \right]\,,
\label{eq:fit_to_cos}
\end{equation}

where $\theta$ is the polar scattering angle between the outgoing 
quark and the incoming electron.

The asymmetry in $\Pep  \Pem \to \PZ \to  \PQb \PAQb$ events is diluted by the effect of mixing in the $\PBz - \PABz$ system. 
Such kind of effect arises from box diagrams involving mainly virtual top quarks~\cite{ALEPH:2005ab,dEnterria:2020cgt}. The average mixing parameter ($\chi$) represents the probability that a produced $\PQb$-hadron decays likewise its antiparticle. The observed $b$-quark asymmetry $\AFBobs$ can then be expressed as $\AFBobs = (1 - 2 \chi ) A_{FB}^{b}$.

%
%
%
%

The combination of LEP measurements at $\sqrt{s} = 91.21-91.26$ GeV resulted in $A_{FB}^{0,b} = 0.0992 \pm 0.0016$, which 
shows today the largest tension in global Electroweak fits 
of the SM~\cite{ALEPH:2005ab,TLEPDesignStudyWorkingGroup:2013myl}. 

A key experimental item for the $A_{FB}^{0,b}$ measurement is the quark \emph{charge} tagging. Among several methods considered at LEP, the lepton-based method identifies the quark charge with the charge of the lepton produced from $\PQb$ or $\PQc$ semileptonic decay~\cite{ALEPH:2001pzx,DELPHI:1994yxx,DELPHI:2003fml,L3:1992fsb,L3:1998bss,OPAL:2003pfe}, while the jet-charge method directly reconstructs the jet charge based on the charges of its tracks~\cite{ALEPH:2001mdb,DELPHI:2004wzo,L3:1998jgx,OPAL:1997goa,OPAL:2002vqi}.

This work studies the $A^b_{FB}$ measurement at the future FCC-ee collider working at the \PZ-boson pole, considering both the lepton-based and jet-charge-based methods and utilizing the official FCC-ee samples and tools.

%% file: sections/mainbody/everything.tex
%
\section{Analysis strategy and event selection}\label{sec:strategy}

%
\subsection{Event generation and detector simulation}\label{sec:mc_samples}
All MC samples are produced with the official FCCSW framework {\footnote{The official FCCAnalysis framework is publicly available as a GitHub repository:  \hyperlink{1}{https://github.com/HEP-FCC/FCCAnalyses}.}} using {\normalfont \scshape Pythia 8}~\cite{Mrenna:2016sih} as generator for the hadronization 
and parton shower processes. The \PQb- and \PQc-quarks fragmentation functions are modelled according to the {\normalfont \ttfamily Lund-Bowler} parameterization~\cite{Andersson:1983ia,Bowler:1981sb}. The IDEA detector~\cite{FCC-Snowmass-2021} response is simulated with the {\normalfont \scshape Delphes 3}~\cite{deFavereau:2013fsa} software using the IDEA detector card~\cite{FCC-ee-IDEA-detector-card-Delphes}.


The signal sample consists of $Z \to \PQb\PAQb$ events, while main background contributions considered are the $Z \to \PQc\PAQc$, $Z \to \PQq\PAQq$\footnote{Here we consider $\PQq=\PQu, \PQd, \PQs$}  and $Z \to \PGmp\PGmm$ processes (always assuming $\sqrt{s}=M_{\PZ}$).
%
%
%



\subsection{Event selection and reconstruction}\label{sec:selection}

\begin{figure}[htb]
\begin{center}
\resizebox{0.50\textwidth}{!}  		
{\includegraphics{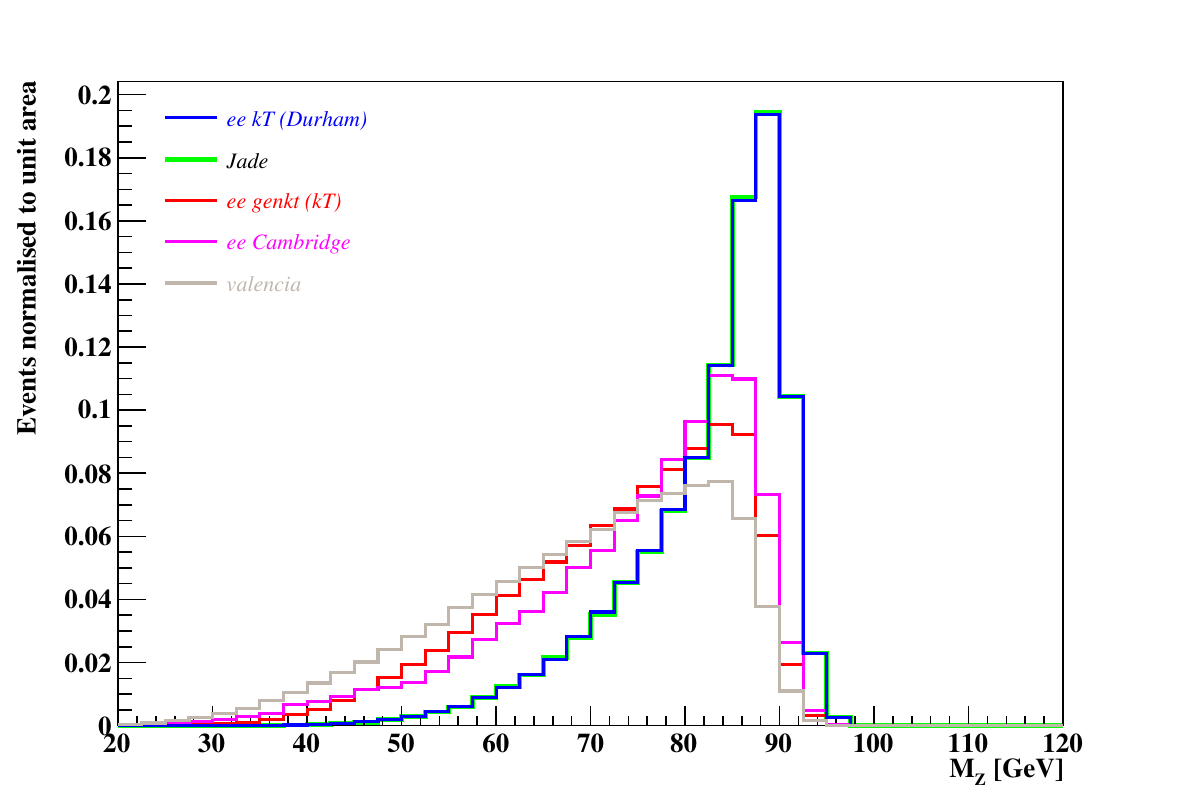}}   
\caption{Reconstructed $Z$-boson mass distribution using different jet clustering algorithms.} 
\label{fig:Z_mass}
\end{center}
\end{figure}

To choose, among all the available jet clustering algorithms in the FCCAnalysis framework, the one which best fits this study, we compare here the {\normalfont \ttfamily Jade}~\cite{JADE:1988xlj,JADE:1986kta} 
and the {\normalfont \ttfamily ee-$k_T$ Durham}~\cite{Catani:1991hj,Lai:2021rko}  
algorithms. For both, we consider the usual {\normalfont \ttfamily E} recombination scheme~\cite{Cacciari:2011ma}, where during the clustering procedure, momenta are combined by simply adding the corresponding four vectors. 
The {\normalfont \ttfamily Jade} algorithm is used as a comparison to be consistent with LEP studies, while {\normalfont \ttfamily ee-$k_T$ Durham} represents a well-developed and up-to-date clustering method. This is used in this study to exclusively cluster events in exactly two jets. Among various jet clustering algorithms shown in \cref{fig:Z_mass}, the {\normalfont \ttfamily Jade} and {\normalfont \ttfamily ee-$k_T$ Durham} ones are as expected the most accurate in reconstructing the $Z$-boson mass distribution, and motivate the usage of the latter throughout the rest of this work. The $Z$-boson mass is here reconstructed as the invariant mass of the two $\PQb$-jets in the final state.

%
%

The following set of selection cuts have been applied. Jets are required to have an energy $E^{\mathrm {jets}} > 10$ GeV and a polar angle $\theta^{\mathrm{jets}} > 0.226$  radians. All events must contain exactly two $\PQb$-tagged jets in the final state. This requirement highly suppresses $Z \to \PQq\PAQq$ and $Z \to \PGmp\PGmm$ background processes. We consider a flat $\PQb$-tagging efficiency of 80\% and a mis-tag rate of 10\% for $\PQc$-quark jets. 
Events are also required to have exactly one positively and one negatively charged $\PQb$-jet in the final state.




\begin{figure*}[htb]
\begin{center}
\resizebox{0.9\textwidth}{!}  		
{\includegraphics{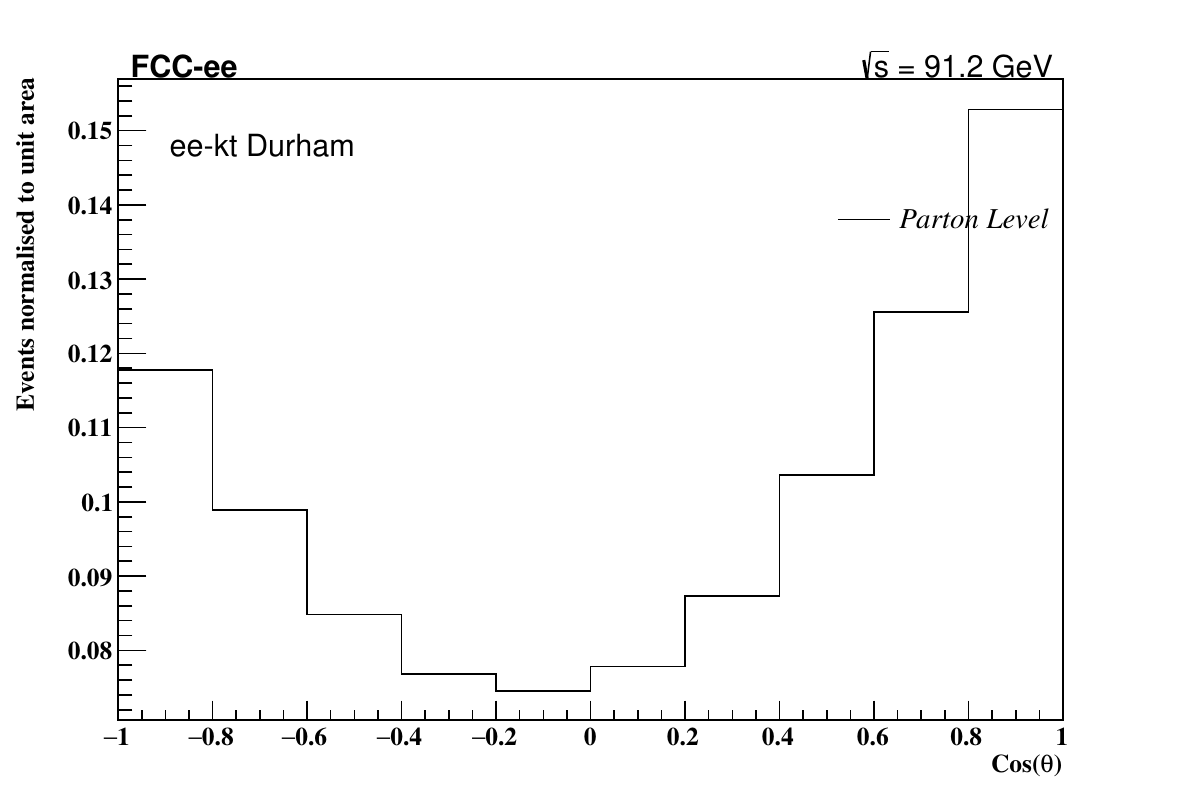}
\includegraphics{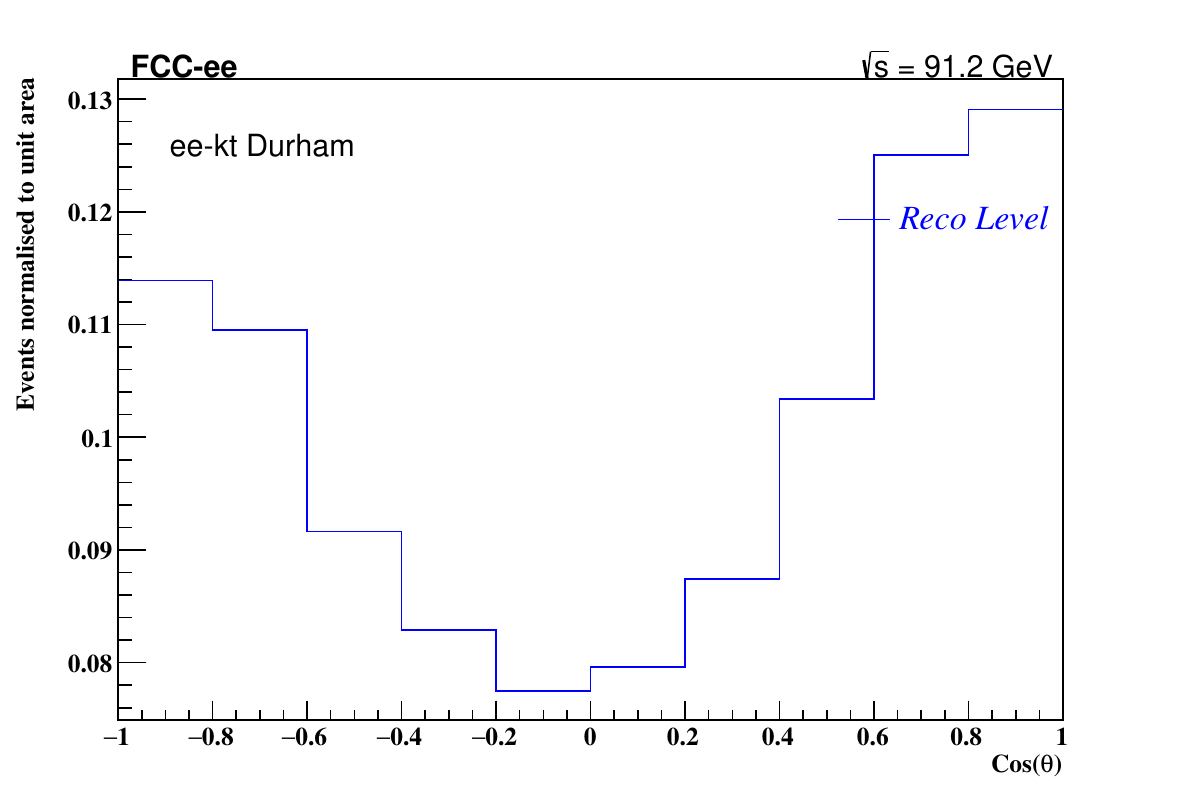}   }   
\caption{Parton level $\PQb$-quark $\cos\theta$ distribution (left) and the corresponding reconstructed $\PQb$-jet $\cos \theta$ distribution (right).}
\label{fig:MC-RECO}
\end{center}
\end{figure*}

\cref{fig:MC-RECO} shows the $\cos\theta$ distribution of the $\PQb$-quark at parton level (left) and the same distribution for the reconstructed $\PQb$-jet (right).

%
\subsection{Lepton-based method}\label{sec:Lepton-based}
In the lepton-based method, the $\PQb$-quark charge is determined from the charge of the lepton produced in the $\PQb$-hadron semileptonic decay. Cascade decays ($\PQb \to \PQc \to \ell$) lead to charge misidentification. To reduce this contribution, the lepton with the highest transverse momentum in each event is considered.

While this method has a relatively low efficiency due to its restriction to semileptonic decays, it typically 
achieves higher purity than the jet-charge-based method.


%
\subsection{Jet-charge-based method}\label{sec:Jet-charge-based}
%

With the jet-charge-based method, the charge of the jet is calculated as the weighted sum of the charges of all the reconstructed tracks associated with the jet:
\begin{equation}
\label{eq:Jet weighted charge}
 Q_{\textup{jet}}=\frac{\sum_i q_i p_{L,i}}{\sum_i p_{L,i}}, 
\end{equation}
where q$_i$ is the $i$-th track's charge, $p_{L,i}$ is the longitudinal momentum of the $i$-th track, the longitudinal direction is defined as the jet direction, and the sum is performed over all the tracks within $\Delta R \leq 0.4$ from the jet\footnote{The angular distance is computed as $\Delta R = \sqrt{\Delta \eta ^2 + \Delta \phi ^2}$, where $\eta$ is the pseudo-rapidity and $\phi$ the azimuthal angle.}. 



%
\subsection{Unfolding and $A_{FB}^{0,b}$  extraction}\label{sec:unfolding}

\begin{figure*}[htb]
\begin{center}
\resizebox{0.9\textwidth}{!}  		
{\includegraphics{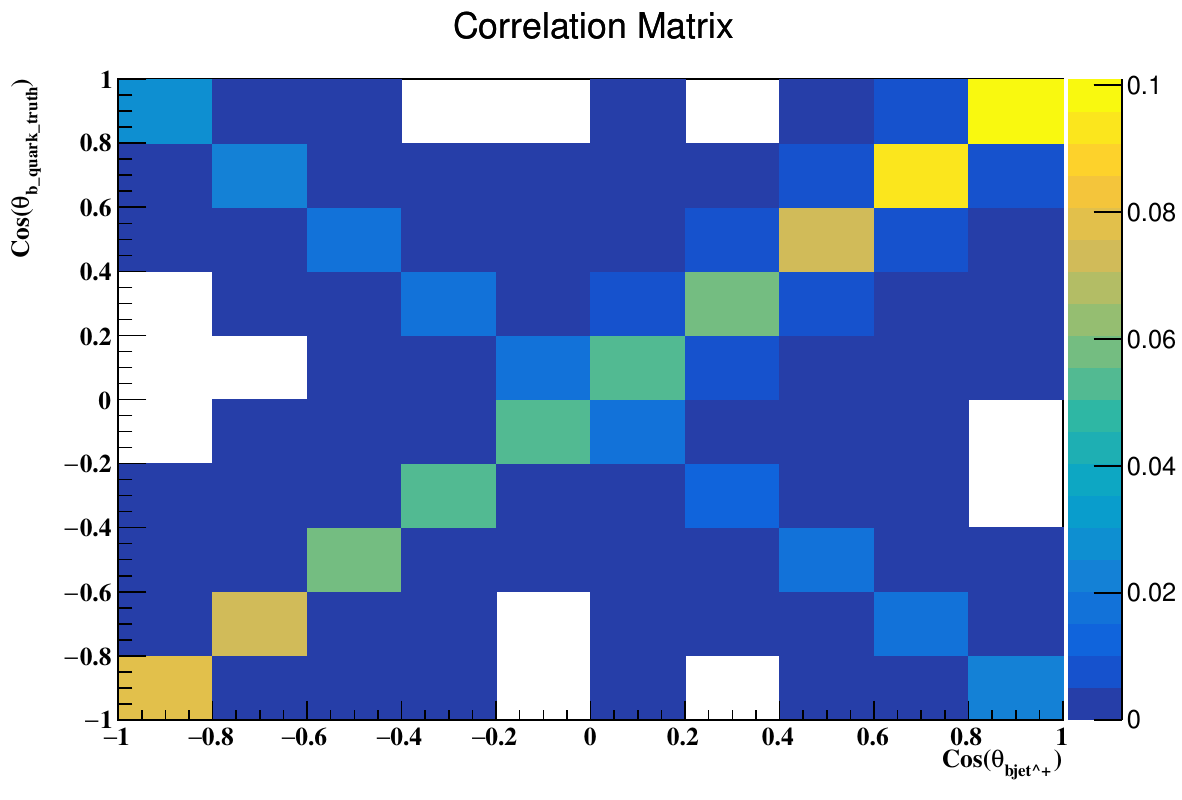}
\includegraphics{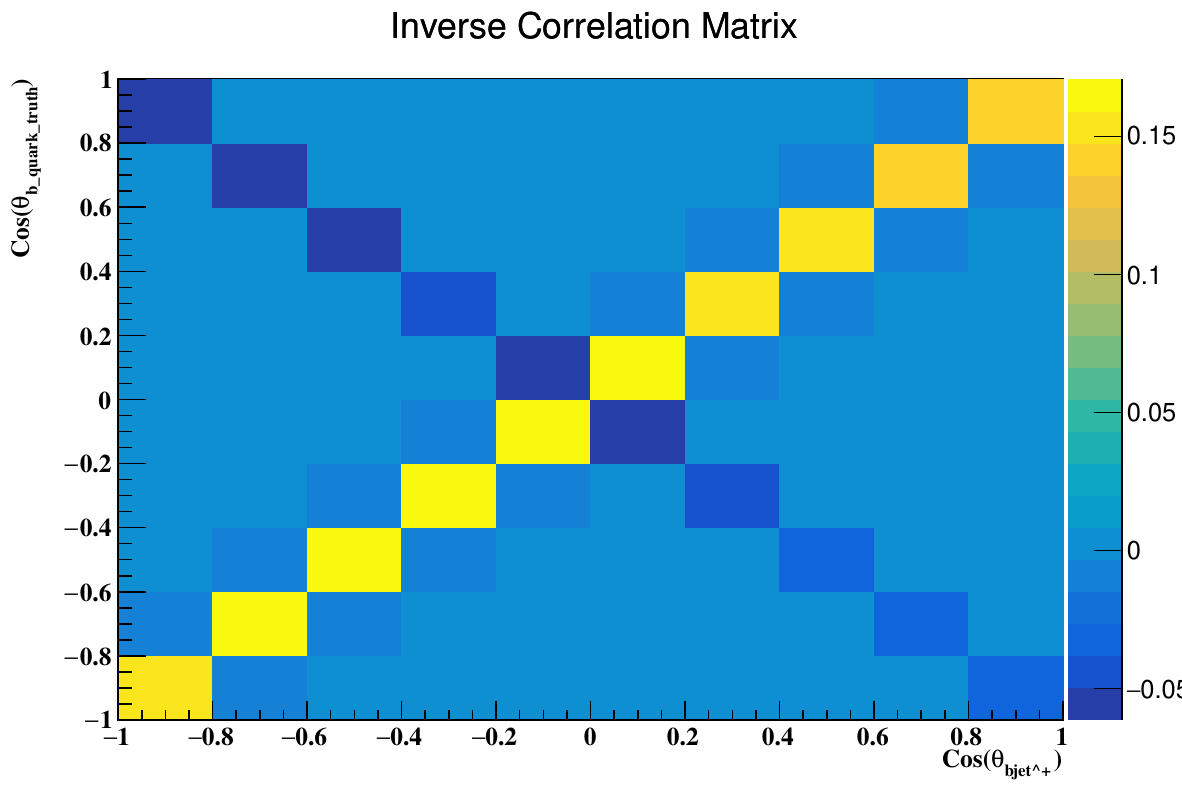}   } 
\caption{Values of the correlation matrix for the ten bins of the obtained differential cross-section and its inverse. } 
\label{fig:Correlation_Matrix}
\end{center}
\end{figure*}

A simplified unfolding procedure~\cite{Choudalakis:2011rr,Biondi:2017pzs} is performed on the obtained differential cross-section with respect to $\cos\theta$. The unfolding correlation matrix and its inverse 
are shown in \cref{fig:Correlation_Matrix}. 
The $A_{FB}^{0,b}$ value has been extracted using \cref{eq:fit_to_cos} by fitting the unfolded differential cross-section~\cite{Tenchini:2008zz}, with ten bins both for the lepton-based and jet-charge-based methods.


%
\section{Statistical and systematic uncertainties}\label{uncertainties}

The reconstruction level differential cross-section distribution is scaled to an expected integrated luminosity of ${\cal L} = 150$ ab$^{-1}$. Statistical fluctuations are created on this distribution via 1000 pseudo-experiments. The unfolding process and the $A_{FB}^{0,b}$ fit extraction are repeated for each replica, and the statistical uncertainty is obtained as the root mean squared (RMS) of the extracted $A_{FB}^{0,b}$ values. The statistical error obtained is~$\sim 10^{-5}$, demonstrating that a precision exceeding that of LEP by two orders of magnitude~\cite{dEnterria:2020cgt}can be achieved.

Different systematic uncertainties have been considered and treated with the standard procedures, and are summarized in \cref{tab:uncertainties}.
\begin{table*}[htbp]  
\begin{center}
\begin{tabular}{l|c|c  }  \hline
Source ~&~ Lepton method  ~&~ Jet-charge method \\  \hline \hline 
b-quark fragmentation  ~&~ -0.00037+0.00091    ~&~  -0.0052+0.0052      \\  
final state QCD radiation ~&~ -0.0034*0.0034   ~&~   -0.00152+0.00152   \\ 
b-tagging efficiency ~&~ -   ~&~   -0.00035+0.00035   \\ 
parton shower modeling ~&~ -0.00012+0.00012   ~&~   -0.00032+0.00032 \\ \hline  \hline
\end{tabular}
\end{center}
\caption{Systematic uncertainties in the $A_{FB}^{0,b}$ measurement.}
\label{tab:uncertainties}
\end{table*}

%
\section{Conclusions}\label{Conclusion}
%

The forward-backward asymmetry of the bottom quark $A_{FB}^{0,b}$  measured around the $Z$-boson pole at LEP remains one of the few experimental 
measurements with a $> 2 \sigma$ discrepancy with respect to the theoretical predictions of the SM.
In this work, the $\PQb$-quark forward-backwards asymmetry measurement has been studied using the lepton-based and jet-charge-based methods, at the FCC-ee collider at $\sqrt{s} = M_Z$ with an integrated luminosity of 150 ab$^{-1}$.
Detailed studies of the statistical and systematical uncertainties have been performed, showing that at the future FCC-ee collider, orders of magnitude improvements with respect to the LEP measurements will be achieved.

The $A_{FB}^b$ value for the FCC-ee at 
the planned integrated luminosity of 150 ab$^{-1}$ at the $Z$-pole is estimated to be
$A_{FB}^b = 0.091170 \pm 0.004907 \, (syst) \pm 0.00001 \, (stat)$ with the lepton-based method and $A_{FB}^b = 0.08866 \pm 0.00296 \, \textup{(syst)} \pm 0.00001 \, \textup{(stat)}$ with the jet charge method.

\newpage